# A novel two- and multi-level binary phase mask design for enhanced depth-of-focus

Vladimir Katkovnik[1], Nicholas Hogasten[2], Karen Egiazarian[1]

[1]Laboratory of Signal Processing, Technology University of Tampere, Tampere, Korkeakoulunkatu 10, FI-33720, Tampere, Finland.

[2]Image Processing Technology, FLIR Systems Inc. 6769 Hollister Ave, Goleta, CA 93117, US.

E-mails: vladimir.katkovnik@tut.fi, nicholas.hogasten@flir.com, karen.egiazarian@tut.fi

**Abstract**

This paper introduces a two-and multi-level binary phase mask design for improved depth of focus. A novel technique is proposed incorporating cubic and generalized cubic wavefront coding (WFC). The obtained system is optical-electronic requiring computational deblurring post-processing, in order to obtain a sharp image from the observed blurred data. A midwave infrared (MWIR) system is simulated showing that this design will produce high quality images even for large amounts of defocus. It is furthermore shown that this technique can be used to design a flat, single optical element, systems where the phase mask performs both the function of focusing and phase modulation. It is demonstrated that in this lensless design the WFC coding components can be omitted and WFC effects are achieved as a result of the proposed algorithm for phase mask design which uses the quadratic phase of the thin refractive lens as the input signal.

## I. INTRODUCTION

Optical systems with diffractive optical elements (DOE) for wavefront coding (WFC) in combination with computational image processing is one of the promising trends in modern optical development. The phase coding results in a blurred image and the final, restored, sharp image is obtained only after digital image processing. It is well known that in systems designed this way not only can the sharpness be restored but also other imaging properties can be improved. In particular, extended depth of focus has been demonstrated using specially designed WFC optics in various applications.

Unlike traditional optics, designed for sharp imaging, system with WFC optical elements produce distorted, in particular blurred, images but the properties of the distortion is such that is allows for efficient image post-processing.

Computational imaging combines novel optical solutions with advanced digital image processing and offers the potential to improve over traditionally designed optics with respect to size, weight, cost and, under some conditions, image quality.

In this paper we design and study two- and multi-level binary phase masks as DOEs. The parameters of the DOE's can be selected in such way that high quality imaging and extended depth of focus (DoF) are achieved without special phase coding elements which are present in conventional WFC designs.

Refractive and diffractive micro-optics developed jointly with advanced digital image processing make it possible to reduce the optics size, weight and cost while making the systems more suitable with respect to the constraints of wafer-level packaging and wafer-level camera fabrication.

Specifically, we study application for thermal infrared imaging in the 2-5 micrometer waveband using a class of DOEs known as Binary Phase Masks (BPM).

The phase mask is called binary if the thickness $h$ takes on a finite number of levels [1]. The mask is truly binary (two-levels) if the number of these different values of $h$ is equal to 2. However, by convention, the term BPM is used for any discrete phase mask with any finite number of levels for $h$. The terms "discrete phase mask" or "multilevel phase mask" are also used.

A phase mask enables a phase delay for an impinging wavefront that is proportional to the thickness of the mask. This phase delay is calculated in radians as

$$\varphi(x,y) = 2\pi \frac{h(x,y)}{\lambda}(n-1), \quad (1)$$

where $h$ stays for thickness of BPM, $\lambda$ is the wavelength and $n$ is the refractive index of the optical material used in the BPM.

The design of a two-dimensional mask is a synthesis of the thickness $h(x,y)$ as a function of the arguments $(x,y)$. Photolithography is one of the relevant technologies for manufacturing BPMs.

We differentiate two types of diffractive optics: *lensless* and *hybrid*.

There is no a refractive lens in the *lensless* setup. It is a pure diffractive optical system. In this case the BPM performs both required functions of focussing and wavefront coding.

The *hybrid* is a refractive-diffractive setup, where *refractive lens* and *diffractive* BPM are combined in a single unit. In the *hybrid* the functions of the lens and BPM are specialized: the refractive lens is responsible for *focusing* and the BPM is responsible for the wavefront *coding*.

*Optimization* is essential in BPM design. The systems assumes joint optimization of the software (image processing algorithm) and the hardware (BPM). In simulation the true synthetic scene is known, thus the accuracy of reconstruction can be calculated. The performance of the imaging system can be characterized by the numerical accuracy of the reconstructed image as compared to the true scene. The design of the BPM is formulated as a constrained optimization. The constraints are defined by the main characteristics of the optical system (PSF, OTF, MTF, etc.) and are averaged over the waveband as well as specified for specific wavelengths as chromatic characteristics of optics. The parameters of the optical setups such as DoF, FoV, aperture size and focal length can be the input parameters of the design or also be included in the constrains.

The paper is organized as follows. In Section II the relevant publications are discussed. Optical setups, notation, modeling and assumptions can be seen in Section III. The design parameters of MWIR optics used in this research are in Section V. The algorithm for the BPM mask design is presented in Section IV. In the subsections V-A and V-B the results for the hybrid and lensless setups are presented. Summary of the results are given in Conclusion. Details of the waveband modelling and of the algorithm proposed for the BPM design are given in Appendix.

## II. Relevant publications

DOEs, e.g. Fresnel lens and BPMs, plus computational inverse imaging represents a technology which is able to relax some of the fundamental restrictions that determine the minimum size, weight and cost of imaging systems using conventional optics. The validity of this statement is confirmed by publications on distinctly different designs of state-of-the-art optical systems, in particular for thermal imagers.

Fresnel lens as a replacement for spherical lens is demonstrated in [2] for LWIR imaging without computational deblurring. Implementation of the cubic and generalized cubic phase mask for WFC and with computational deblurring can be seen: for visual waveband in [3], [4], for low-infrared (LWIR) in [5], for near-infrared (NIR) with athermalization in [6], for LWIR with athermalization in [7]. Application of generalized cubic phase mask for design of zoom systems for visual waveband is a topic of papers [8], [9], [10], [11], [12]. A wide variety of different types of phase masks have been studied in order to increase the depth of field: cubic and generalized cubic [3], [4] logarithmic [13], exponential [14], hyperbolic [15], square-root [16], etc.

Some of these masks have been compared in [17], where signal-to-noise ratio is taken as a comparative metric. Optimization of WFC based on the phase transfer function is proposed in [18]. The importance of WFC, review and further results are discussed in [19], [20] (see also [21]).

Spatial light modulator (SLM) implementations of this kind of WFC for improved DoF is demonstrated in [22], [23].

BPMs is a special branch of development and application of DOEs. First of all we need to mention the fundamental publications [1] and [24]. Most publications on BPM design (two- and multi-levels) are restricted to symmetric masks composed from concentric rings enabling different phase delay for the wavefront forming [25], [26], [27].

To the best of our knowledge we have not seen approaches incorporating the specific WFC properties in design of BPMs which is one of the results of this paper.

The term "optical-electronic imaging" is commonly used for the system requiring post-processing to reconstruct a sharp image from a deliberately distorted/coded blurred observed/captured image. The deblurring algorithm is a sub-element of the system design because the final equality depends on both the optics and the algorithm used. For instance, in [28] a total variation regularized deconvolution algorithm was developed for noisy Poissonian observation. The comparison with the essential advantage of the proposed algorithm is produced versus the Wiener filter (e.g. [29]) conventional for this sort of problems.

In this paper we use Block Matching 3D Deblurring (BM3D-DEB) algorithm [30] which is one of the most successful and universally applicable for imaging application. The algorithm consists of two stages: first, Wiener inverse imaging, and second BM3D filtering. The success of the algorithm is enabled mainly by this filters use of a flexible and adaptive sparse modeling of the images to be reconstructed [31]-[33].

## III. OPTICAL SETUP, NOTATION, ASSUMPTIONS

We consider a singlet lens optical model depicted in Fig.1: the object plane with coordinates $(\xi, \eta)$, the lens plane $(x, y)$ and the sensor array plane $(u, v)$. It is assumed that the object and lens are thin, $z_1$ is a distance from the the object to the lens, $z_2$ is a distance from the lens to the sensor (image) plane, and $f$ is a foc

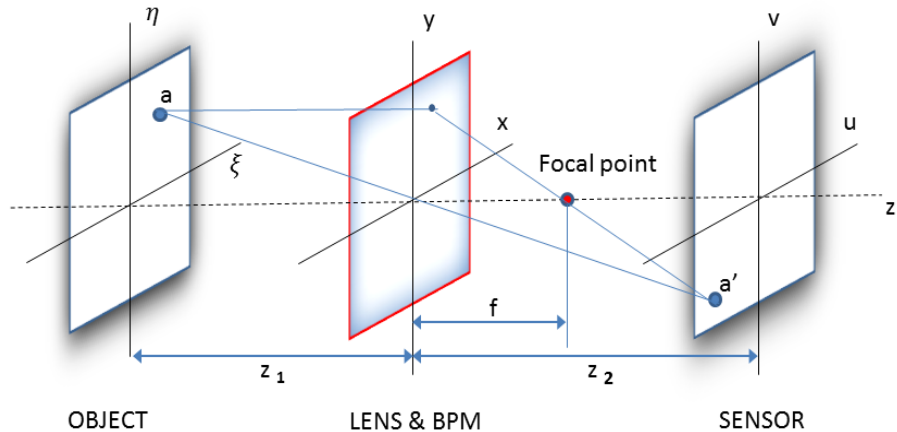

Fig. 1. Singlet optical setup with lens & BPM in the pupil plane.

This setup is standard for studies on DOE design using also the conventional assumptions that all optical elements are thin and located at the lens plane. Under a paraxial approximation and Fresnel optics assumptions the generalized pupil function of the considered system is of the form:

$$P_g(x,y) = P_A(x,y) \exp\left[\frac{j\pi}{\lambda}\left(\frac{1}{z_1} + \frac{1}{z_2} - \frac{1}{f}\right)(x^2 + y^2) + jBPM_{\lambda_0,\lambda}(x,y)\right]. \qquad (2)$$

Here $\lambda$ is the wavelength, $P_A(x,y)$ is the aperture of the lens and $BPM_{\lambda_0,\lambda}(x,y)$ represents the phase delay enabled by BPM for this wavelength provided that $\lambda_0$ the wavelength parameter is used for BPM design.

The normalized *point spread function* (PSF) for the incoherent version of the system shown in Fig. 1 accordingly to [34] is of the form

$$PSF(u,v) = |\mathcal{F}_{P_g}(\frac{u}{z_2\lambda}, \frac{v}{z_2\lambda})|^2 / \iint_{-\infty}^{\infty} |\mathcal{F}_{P_g}(\frac{u}{z_2\lambda}, \frac{v}{z_2\lambda})|^2 du dv, \tag{3}$$

where $\mathcal{F}_{P_g}$ stands the Fourier transform of $P_g$.

The the *optical transfer function* (OTF) can be written as

$$OTF(f_X, f_Y) = \iint_{-\infty}^{\infty} PSF(u,v) \exp[-j2\pi(f_X u + f_y v)] du dv. \tag{4}$$

Let $I_s(u,v)$ and $I_o(u,v)$ be the intensities of the wavefronts (images) at the sensor and object planes, respectively.

In our simulation experiments, observations $z(u,v)$ are obtained by convolving the true image $I_o(u,v)$ with $PSF(u,v)$ of the system. To make the data more realistic for a true imaging system we add a white Gaussian noise:

$$z(u,v) = \int_{-\infty}^{\infty}\int_{-\infty}^{\infty} PSF(u-u', v-v') I_o(u', v') du' dv' + \varepsilon(u,v), \tag{5}$$

and in the Fourier domain this gives

$$z(f_X, f_Y) = OTF(f_X, f_Y) I_o(f_X, f_Y) + \varepsilon(f_X, f_Y)$$

where $\varepsilon(u,v) \sim N(0, \sigma^2)$.

Recall, that in these formulas the object $I_o(u,v)$ is resized to the image sensor size and thus the magnification factor $M = -z_2/z_1$ is eliminated.

By convention the lens defocus in (2) is characterized by the parameter

$$\Psi = \frac{\pi}{\lambda}\left(\frac{1}{z_1} + \frac{1}{z_2} - \frac{1}{f}\right)(D/2)^2, \tag{6}$$

where $D$ is the lens diameter.

This $\Psi$ is the maximal value of the phase $\frac{\pi}{\lambda}\left(\frac{1}{z_1} + \frac{1}{z_2} - \frac{1}{f}\right)(x^2 + y^2)$ in radians caused by the defocus. Let the focus be ideal with the distances $z_{1,0}$ and $z_{2,0}$

$$\frac{1}{z_{1,0}} + \frac{1}{z_{2,0}} - \frac{1}{f} = 0$$

and the deviation from the best focus image plane position be defined as $\Delta z_2 = z_2 - z_{2,0}$. Then for small $\Delta z_2$ we obtain

$$\Psi \simeq -\frac{\pi}{\lambda z_{2,0}}(D/2)^2 \cdot \frac{\Delta z_2}{z_{2,0}} \tag{7}$$

In what follows we characterize the defocus by the ratio $\Delta z_2/z_{2,0}$, i.e. by a relative deviation of the image plane from the best focus position. In this notation, the maximum value of $|\Delta z_2|$ still allowing high quality of imaging defines DoF.

In our experiments $|\frac{\Delta z_2}{z_{2,0}}|$ is restricted by $0.04$. For the parameters in our experiments it corresponds to variations in $\Psi$ such that $|\Psi| \leq 30$ radians, which still results in noticeable defocus.

Roughly speaking, there are two approaches to BPM design for DoF enhancement. In the first approach the optical characteristics of the system such as $PSF$ or $OTF$ are optimized with the main intention of getting the optical system robust with respect to $\Psi$ and to avoid values close to zero for $OTF$ in the image effective bandwidth. In the second approach the system is considered from end-to-end, i.e. from observation to reconstruction including the used image processing algorithm. The main criteria of concern is the accuracy of imaging and the optical characteristics of the system are treated as an auxiliary parameters [35], [17].

The method considered in this paper we are based on the second approach. One of our reasons for selecting this path is the use of post-processing, the BM3D-DEB algorithm. This algorithm is based on sparse representation of the image signal to be reconstructed and to some extend it is robust to the effects of $OTF$ values close to zero. For evaluation the feasibility of a BPM system we use the final imaging results, post signal processing.

The reconstruction of $I_o$ from observations $z$ is an inverse imaging (deblurring) problem. The efficiency of this inverse imaging will be evaluated in terms of pick-signal-to-noise-ratio (PSNR) measured in dB:

$$PSNR = 20\log \frac{\max_{u,v} I_o(u,v)}{RMSE} \ (dB), \qquad (8)$$
$$RMSE = \sqrt{mean_{u,v}([\hat{I}_o(u,v) - I_o(u,v)]^2)},$$

where $\hat{I}_o$ is a reconstruction of $I_o$, $\max_{u,v} I_o(u,v)$ is a maximum (peak value) of the object image intensity and the Root-Mean-Square Error ($RMSE$) calculated as the square-root of the mean value of the squared error between the true image $I_o(u,v)$ and its reconstruction $\hat{I}_o(u,v)$ obtained after deblurring.

The optimization is produced with respect to the BPM by maximizing a criterion which can be formalized as follows

$$J = \max_{\Psi} PSNR \text{ provided } min_{\Psi}PSNR > 30. \qquad (9)$$

The maximum $PSNR$ is usually achieved at the focal point. Thus maximizing $J$, we maximize the quality of the image with respect to focus, provided that for any defocus $|\frac{\Delta z_2}{z_{2,0}}| \leq 0.04$, $PSNR$ is larger than 30 dB. It is generally accepted that $PSNR$ values higher that 30 $dB$ result in a high quality image. Of course even larger value can be taken for this low bound of $min_{\Psi}PSNR$ but there is a risk that there is no solution for (9).

The BPM component in (2) depends on the design wavelength $\lambda_o$ and the image wavelengths $\lambda$. The BPM in this paper is designed for mid-wavelength infrared (MWIR) imaging within a given waveband.

## IV. Design of BPM

Phase modulation with $BPM_{\lambda_0,\lambda}$ in the generalized pupil function Eq.(2) is used in order to extend DoF. The basic idea behind the used approach, known as WFC, is to change the focusing properties of the imaging system such that the $OTF$ is almost invariant with respect to the defocus $\Psi$ whish allows us to use this $OTF$ for inverse imaging for all amounts of defocus.

Starting from the seminal works by Dowski, E. R. and Cathey, W. T. [4] and [3] there has been a flow of publications on phase modulation design, analysis and optimization for imaging based on digital post-processing.

While the initial motivation for WFC was an extended DoF, it has been demonstrated that WFC is an universal instrument for design which is robust with respect to various types of disturbances and aberrations, e.g. chromatic aberrations, errors in optics manufacturing and assembly, athermalization, etc.

The cubic phase mask (CPM) and the generalized cubic phase mask (GCPM) are among the most popular. They are defined by the formulas:
- CPM:
$$\varphi(x,y) = \alpha(x^3 + y^3);$$
- GCPM:
$$\varphi(x,y) = \alpha(x^3 + y^3) + \beta(x^2 y + y^2 x). \qquad (10)$$

Beyond the cubic phase masks, various candidates for WFC have been studied: exponential, square-root, logarithmic, harmonic, etc. (e.g. [15], [16]). Note, that the functions $\varphi(x,y)$ used for WFC are not restricted to the interval $[0, 2\pi]$, i.e. they define absolute phases.

While the various phase masks yield different effects, qualitatively, the achieved imaging results are quite similar. Thus, GCPM can be treated as an universally applicable model for WFC.

The binary phase masks (BPM) is a special class of phase manipulators. It is a thin diffractive plate. The term binary is equally used for true binary (two level) and multilevel masks. BPMs are used for the phase modulation of the wavefront in the camera aperture. The goal of this modulation can vary from minor camera improvement, for example through compensation of aberrations, to full replacement of potentially bulky and expensive refractive lenses. Recently, BPMs have become a popular tool for design of "lensless" micro-cameras. The BPMs are usually designed as compositions of concentric rings of different width and radius and the design focuses on optimization the selection of these geometric parameters.

Our approach to the BPM design is very different. A continuous absolute phase $\varphi(x,y)$, say CPM or GCPM, is used as an input variable and the binary mask is obtained as a special nonlinear transformation of this phase function.

The design algorithm is composed of the following three successive stages:

(1) A special, piece-wise invariant, approximation of $\varphi(x,y)$ as $\hat{\varphi}(x,y)$ with a prescribed value for the minimum size of the invariant area of the mask defined by the parameter $m_w$. It can be called *step-width parameter as* $\hat{\varphi}(x,y)$ is a step-wise function.

(2) Wrapping of $\hat{\varphi}(x,y)$ to the desired upper bound of the mask width (range) $\pi m_0$ in radians, i.e.

$$|\hat{\varphi}(x,y)| \leq \pi m_0. \tag{11}$$

We use the term *order* for the parameter $m_0$.

(3) Discretization of this wrapped phase to a prescribed maximum number of the levels $N$.

Three parameters control the results of this design: the minimal width of the piece-wise invariant segments calculated as $m_w - 1$, the phase range of the mask defined by the order $m_0$ and the maximal number of levels $N$. These parameters are the variables of optimization in (9).

We consider two types of optical systems: hybrid (lens & BPM) and lensless (only BPM).

In the hybrid system the PSF is defined by (2), i.e. the lens is responsible for focusing and WFC is accomplished by the BPM. In order to design the BPM we use the phase function $\varphi(x,y)$, e.g. (10), selected as an input variable of the above algorithm.

In the lensless system there is no lens and both operations, focusing and wavefront coding, are performed by the BPM. For this case (2) takes the form

$$P_g(x,y) = P_A(x,y) \exp\left[\frac{j\pi}{\lambda}\left(\frac{1}{z_1} + \frac{1}{z_2}\right)(x^2 + y^2) + jBPM_{\lambda_0,\lambda}(x,y)\right]. \tag{12}$$

Comparing this formula with (2) we may conclude that the phase to be coded in BPM is defined as

$$\varphi(x,y) = -\frac{\pi}{\lambda f}(x^2 + y^2) + \varphi_i(x,y), \tag{13}$$

where $-\frac{\pi}{\lambda f}(x^2 + y^2)$ is the phase shift corresponding to the omitted lens and $\varphi_i(x,y)$ is targeted on WFC.

The BPM for (12) is designed by the above algorithm with the absolute phase (13) as an input variable.

Despite the additive form of the focusing and WFC components in (12) these function are not additive in BPM and cannot be separated.

## V. MWIR IMAGING

The focal length, aperture diameter, field of view (FoV) and depth of focus (DoF) are the first order parameters of an optical solution. Based on these parameters we model and analyze optical setups with a generalized aperture equipped with DOE.

The proposed approach is applied for thermal infrared imaging. In what follows we report the results of a hypothetical MWIR system with typical optical parameters. The used parameter values are shown in Table I.

TABLE I
PARAMETERS OF MWIR SYSTEM

| $HFoV$ | $F\#$ | $D_0$ | $f_0$ | Waveband | Pixel pitch | Resolution |
|---|---|---|---|---|---|---|
|  |  | $mm$ | $mm$ | $\mu m$ | $\mu m$ |  |
| $10°$ | 2 | 4.6 | 9.1 | 3.4-5.1 | 5 | 1280×1024 |

Here $HFoV$ is a horizontal field of view, $D_o$ is a lens diameter, $f_o$ is a lens focal distance. The parameters in the two last columns define the sensor pixel pitch and the sensor resolution defined as [columns × rows].

The $PSF$ is normalized for the given effective waveband using a weighted average of the PSF (2) for every wavelength in the waveband. The weights are defined by the Planck curve for black-body radiation (details in Appendix B). In a final system design the spectral transmission properties of the lens and/or BPM material and coatings as well as the spectral absorption properties of the sensor would need to be taken into account. In this paper we assume those properties are constant across the effective waveband.

### A. Hybrid optical system

Let us show the results for the hybrid optics with the following BPM parameters : the order $m_0 = 1$, the number of levels $N = 4$, the width-parameter $m_w = 16$, $\alpha = 6.5$, $\beta = 0$. The order $m_0$ defines the upper bound for the thickness (range) of the mask in radians. According to Eq.(11), it means that for $m_0 = 1$ the difference between the largest and smallest values of the phase in BPM is smaller or equal to $2\pi$. In what follows we assume that $z_1 = 100$ m and $z_2 = 0.0091$ m.

In Fig.2 we show $2D$ image of the designed BPM and its cross-section. The $(x, y)$ coordinates in the 2D image are given in wavelengths. The phase values of BPM are in radians. The cross-section provides an explicit information about a $3D$ structure of the BPM. First, we may note that the thickness (range) of BPM is less than $2\pi$, what correspond to the order $m_0 = 1$. The number of levels of the BPM is equal to $N = 4$. The cross-section is a step-wise invariant function. The width-parameter $m_w = 16$ means that the minimum size (width) of the horizontal steps of the cross-section is calculated as $m_w - 1 = 15$. The designed BPM is a piece-wise invariant 3D surface with quite large horizontal patches in $(x, y)$ directions having sizes not less than $m_w - 1 = 15$.

The cross-section demonstrates that the profile of the mask is quite simple and reasonable to implement using conventional etching or laser lithography techniques.

The comparative performance of the hybrid system is illustrated in Fig.3. It is done for noisy data with $\sigma = 0.002$ in (5), i.e. about $SNR = 48$ dB with respect to an additive zero-mean Gaussian noise in the sampled blurred image. The result for the hybrid solution (lens & BPM) is shown for unprocessed observations dash-circle (red) and the deblurred reconstruction solid-circle (red). The equivalent results are shown for the optical lens system with no BPM, observations dash-cross (green) and deblurred reconstruction solid-cross (green) and for the system with optical lens plus continuous CPM, observations dash-diamond (black) and reconstruction solid-diamond (black).

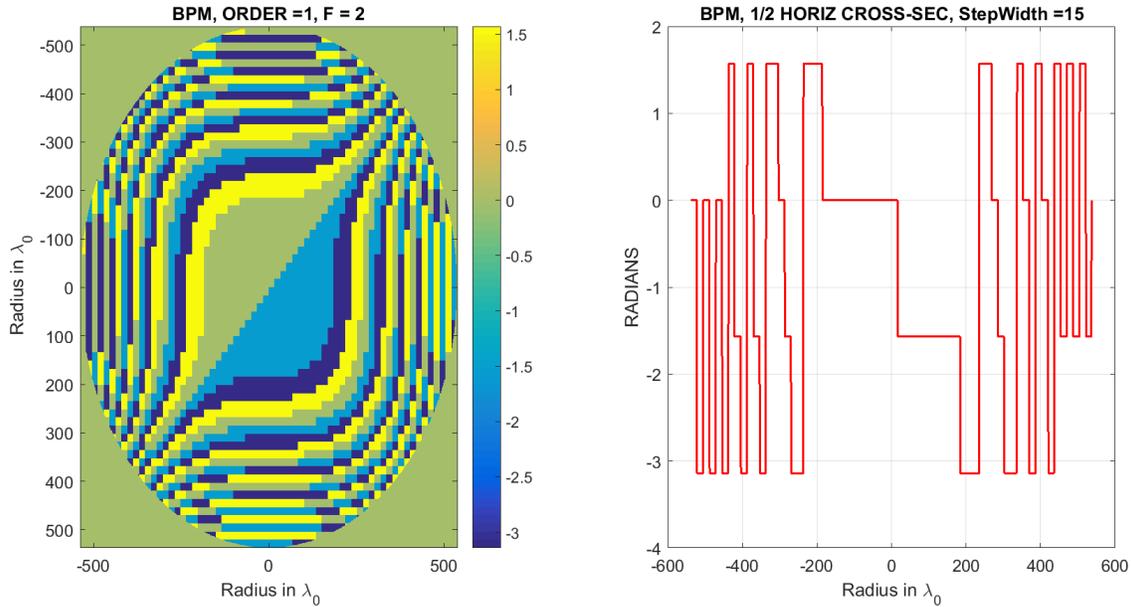

Fig. 2. Hybrid: lens & BPM, order $m_0 = 1$, level number $N = 4$, $m_w = 16$, $\alpha = 6.5$, $\beta = 0$.

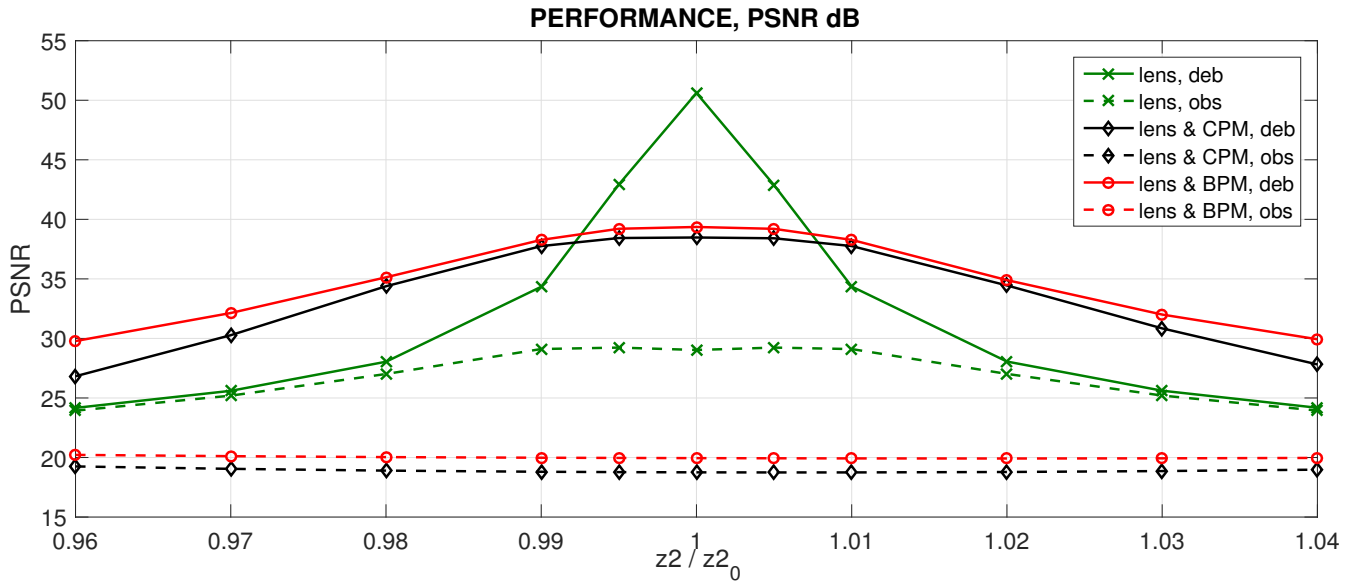

Fig. 3. The optical lens system with no BPM, observations dash-cross (green) and deblurred reconstruction solid-cross (green). The optical lens with continuous CPM system, observations dash-square (black) and reconstruction solid-square (black). The result for the hybrid with BPM are shown as observations dash-circle (red) and deblurred reconstruction solid-circle (red).

The optical lens without any phase modulation will have improved performance with deblurring, but only in the very narrow neighborhood of the focal distance with deviations less that $1\%$ of the focal distance. As expected the two systems with BPM and CPM show a large improvement in PSNR after deblurring and the improvement stays significant for relatively large deviations from ideal focus of up to $4\%$ of the focal distance.

It is interesting to observe that for this simulation the discretized BPM actually outperforms its continuous CPM equivalent. If we treat the BPM as a discrete approximation of continuous CPM, it is a very useful approximation because it performs better than its original.

The comparison of the solid and dash curves demonstrates how efficient the BM3D-DEB technique is;

even for the lens system with no phase modulation.

The visual comparison of the results for the strongest $4\%$ misfocus is shown in Figs.4. The left image pair shows the observed and deblurred result from the hybrid optical system and the right image pair shows the observed and deblurred result from the lens-only system. The improvement in sharpness of the hybrid solution is clear. The plots below show a line plot of the ground truth signal (solid red) for the center line and the observed signal (dot-dash black) pre and post reconstruction.

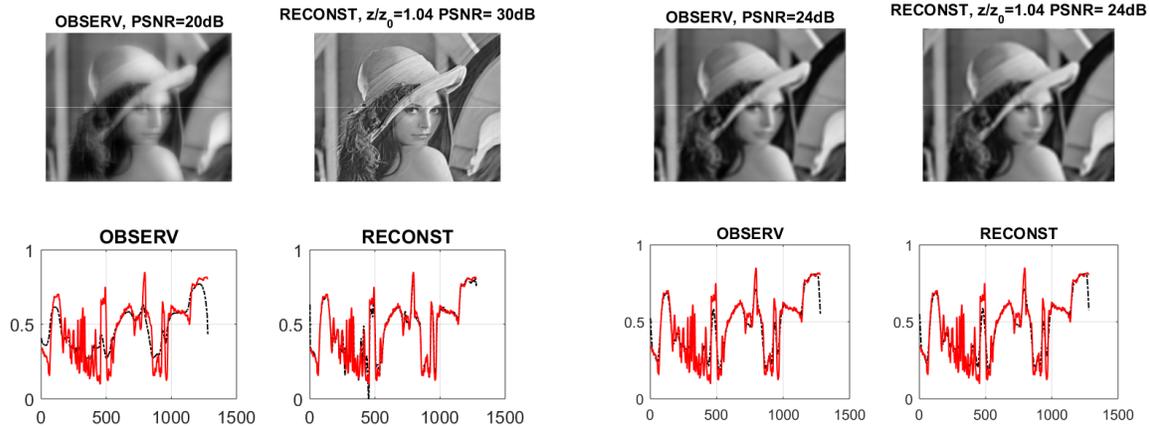

Fig. 4. The left and write images are done for the hybrid and lens only systems, respectively, provided the strong 4% misfocus. The advantage of the reconstruction obtained for the system with the BPM phase manipulation is clear.

The longitudinal cross-sections of PSF's produced through the focal point are shown in Figs.5. The left image is for the lens system without phase modulation. It is very sharp with the peak located exactly at the focal point. The right shows the PSF for the hybrid solution with BPM. It is much wider and elongates along the axis $z$, which defines the robustness of the system with BPM with respect to the misfocus.

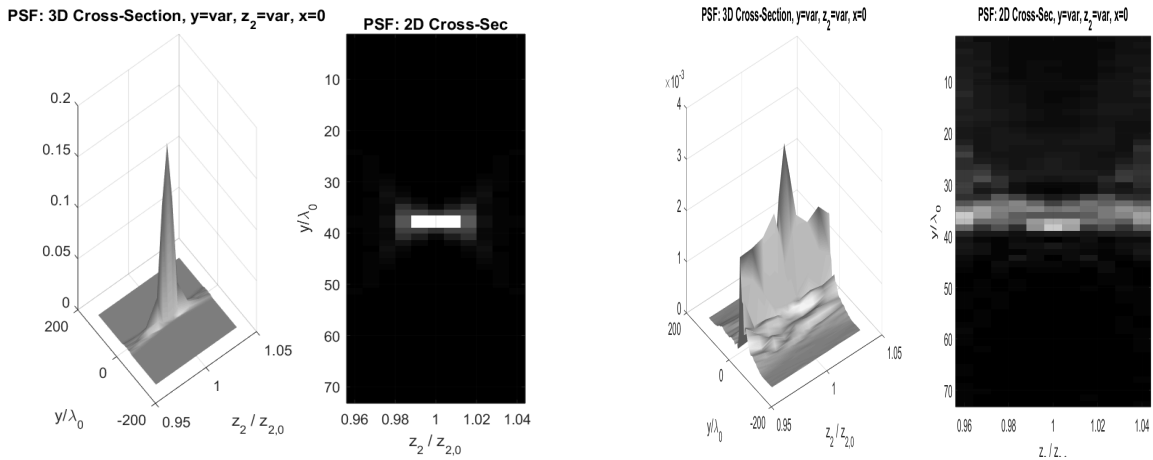

Fig. 5. The longitudinal cross-sections of PSF's for the lens system without any phase modulation and with BPM. For the lens (left image) it is very sharp with the pick located exactly at the focus point. The right image for hybrid with BPM mask is wide and elongated along the axis z, what defines a much less sensitivity of the system with BPM with respect to the defocus.

Images in Figs.6 and 7 show the dependence of MTF on variations in the defocus $z$ and the wavelength $\lambda$. Plots on the left are prepared for the lens-only system without phase modulation and plots on the right show results for the hybrid system with BPM. In each plot we have a set of curves. In Fig.6 the effective waveband is fixed in each sub-plot and curves are produced for a set of $z$ covering the interval of defocus. In Fig.7 the amount of defocus is fixed in each sub-plot and curves are produced for a set of varying $\lambda$ covering the effective waveband interval. The general conclusion is obvious, while the MTF for the lens system is very sensitive with respect to both the defocus and change in wavelength (color), the BPM modulation results in strong stabilization of the MTF which becomes robust with respect to any variations in either defocus and spectral characteristics of the signals.

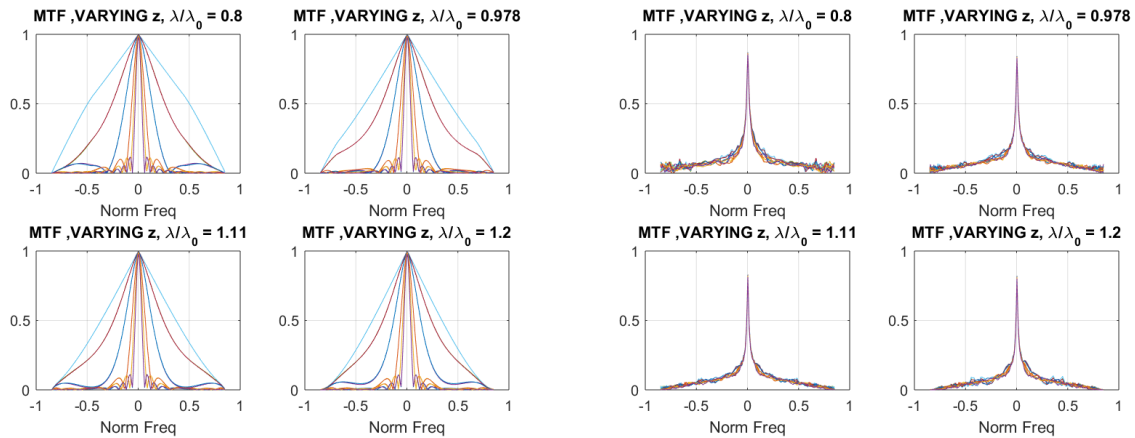

Fig. 6. The dependencies of MTF on variations in the defocus z and the wavelength. Left images are done for the lens system without phase modulation and right images for the system with BPM. In each sub-image we have a set of various curves. The wavelength is fixed while the set curves are obtained for various defocus $z$.

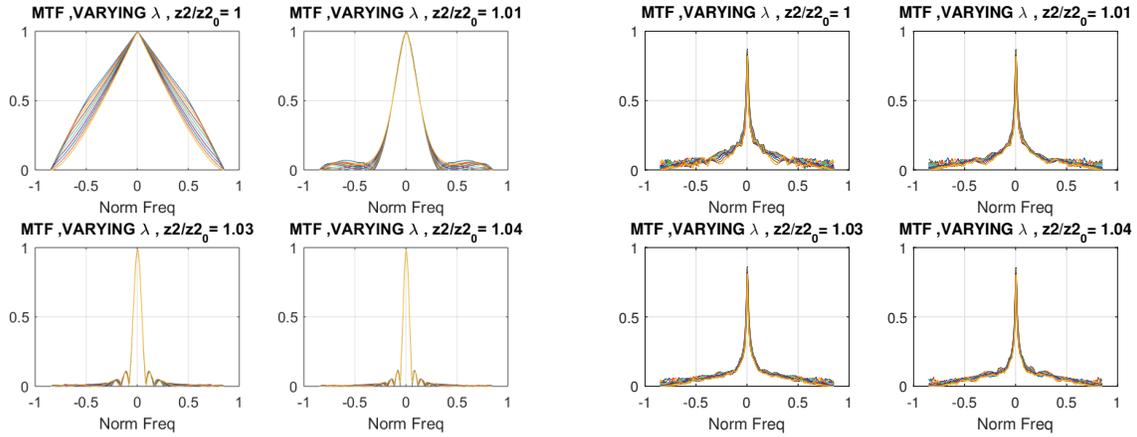

Fig. 7. The dependencies of MTF on variations in the defocus $z$ and the wavelength. Left images are done for the lens system without phase modulation and right image for the system with BPM. In each image we have a set of various curves. The defocus z is fixed while the set curves are obtained for various wavelength.

## *B. Lensless BPM optical system*

In the lensless optical setup the BPM is responsible for both focusing and wavefront coding, i.e. the both cubic and quadratic phase variations are exploited for the mask design.

Surprisingly the specific WFC effects have a place even without the cubic term $\varphi_i(x,y)$ in (13), i.e. the system with this sort of BPM, designed for pure quadratic phase, demonstrates decreased sensitivity with respect to defocus.

In Figs.8 and 9 we show the two and four levels BPM designed using the WFC's cubic and lens's quadratic phase. We considered also the designs without the cubic summands in (13). Visually these masks are very similar to those shown in Figs.8 and 9.

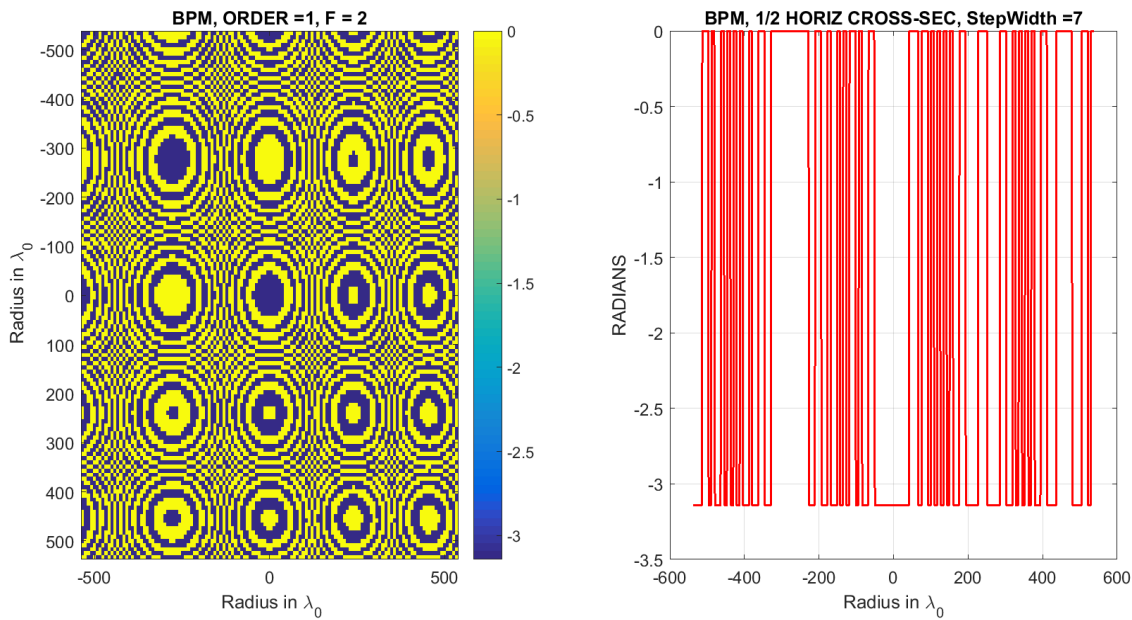

Fig. 8. Two level BPM for the lensless system.

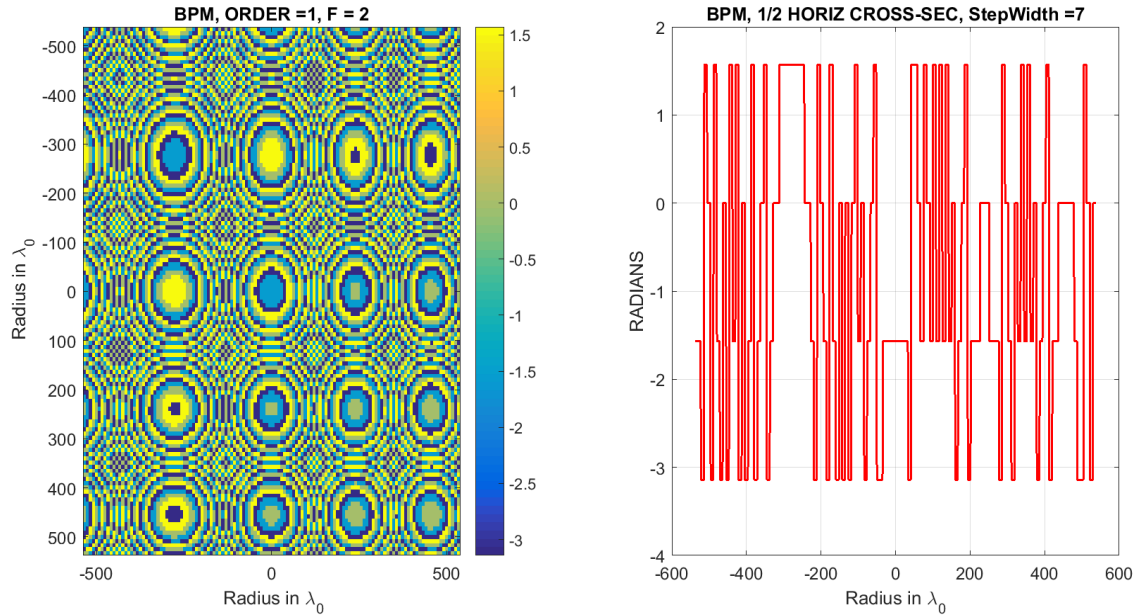

Fig. 9. Four level BPM for the lensless system.

The performance of these BPMs is shown in Fig.10. The four curves on the top corresponds to the PSNR of deblurred images. The two solid (red) curves correspond to the BPM with the cubic phase terms while the other two dashed (black) correspond to the pure quadratic phase coded in BPM.

Firstly we note that all these curves are nearly flat with a high level of PSNR (larger than 30 dB desirable in (9)) which is an indication that the final, deblurred, image quality is very good. The curves corresponding to the pure quadratic phase differ only by a low magnitude with respect to the more stable flat curves obtained for the phase coding with cubic components. The four levels BPMs show about 5 dB better performance than the two level BPMs.

The group of the four horizontal curves (two of them directly overlaid and not individually distinguishable) at the bottom of this figure correspond to the PSNR of the noisy observations and show how much the deblurring is able to improve imaging.

The longitudinal cross-sections of PSF's for the lensless systems without cubic and with cubic phase terms are shown in Fig.11. Comparison of these curves versus the similar curves for the lens system in Fig.5 demonstrates the WFC effects in the lensless case and shows that the cubic term makes this cross-section more uniform with respect to $z$.

In Fig.12 we show the variations of MTF of the lensless, two level, binary system with respect to the defocus $z$ and the wavelength $\lambda$. The comparison with Figs.6 -7 demonstrates that the stabilization effects observed in the hybrid system is not as strong in the lensless case.

Nevertheless the PSNR curves in Fig.10 show that, averaging over the effective waveband, the imaging enabled by the lensless system with BPM is of high quality.

## VI. Conclusion

A novel BPM based technique, designed for improved depth of focus, is proposed, allowing the incorporation of WFC. The obtained system is optical-electronic requiring computational deblurring post-processing in order to obtain a sharp, high quality image from the observed poor signal. It is shown in the simulation of MWIR imaging that this design is quite efficient producing high quality imaging even for high levels of misfocus. It is also shown that this technique can be used for design of a lensless system. It is demonstrated that in the lensless design WFC coding components can be omitted and desired

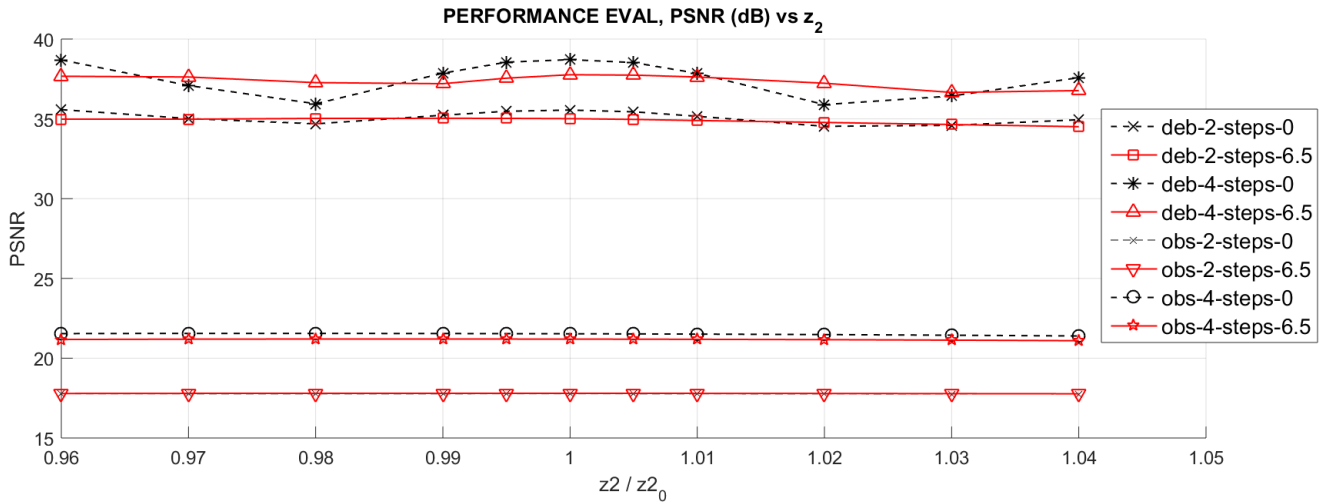

Fig. 10. The two solid (red) curves at the top corresponds to BPM with the cubic phase terms while the other two dash (black) correspond to the pure quadratic phase coding in BPM. The four curves at the bottom correspond to observation PSNRs. Details on correspondence of the curves to different cases are clarified by the legend.

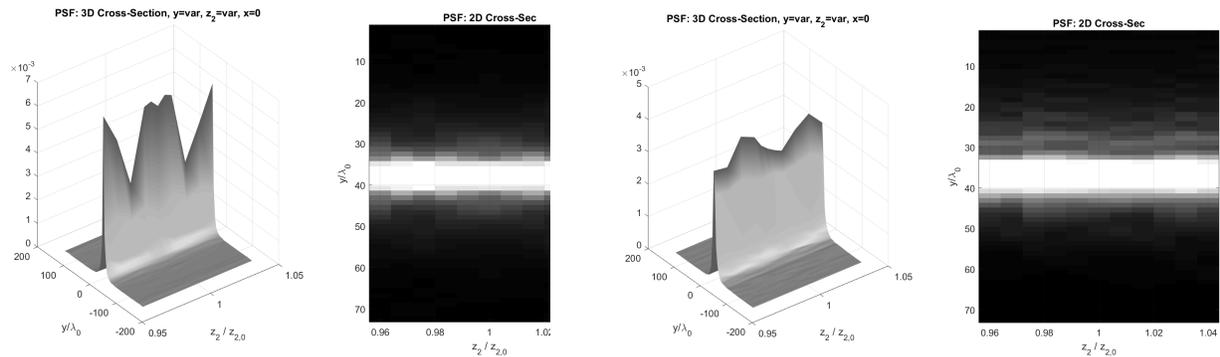

Fig. 11. The longitudinal cross-sections of PSF's for the lensless system without cubic (left) and with cubic component (right) in BPM design.

WFC effects will still be present as a result of the proposed algorithm for BPM design using as an input signal the quadratic phase of the thin refractive lens. Its efficiency is demonstrated for thermal imaging wavebands for two configurations of the system: lensless (only BPM) and hybrid (lens & BPM).

The algorithm developed for BPM design is universal and can be applied for various scenarios not restricted to MWIR imaging and for various phase functions enabling WFC.

## VII. Acknowledgement

This work was supported by Academy of Finland, project no. 287150, 2015-2019.

## Appendix

### A. Waveband modeling

The phase shift by BPM in (2) is defined by the function $BPM_{\lambda_0,\lambda}(x,y)$ depending on the wavelength $\lambda$ and the wavelength $\lambda_0$, which serves a design parameter.

The $PSF$ depends on both $\lambda$ and $\lambda_0$. Denote it as $PSF_{\lambda,\lambda_0}(u,v)$.

For a non-monochromatic radiation the intensity at the sensor plane is calculated as

$$I_s(u,v) = \sum_\lambda \iint_{-\infty}^{\infty} PSF_{\lambda,\lambda_0}(u-u',v-v')\ I_{o,\lambda}(u',v')du'dv', \qquad (14)$$

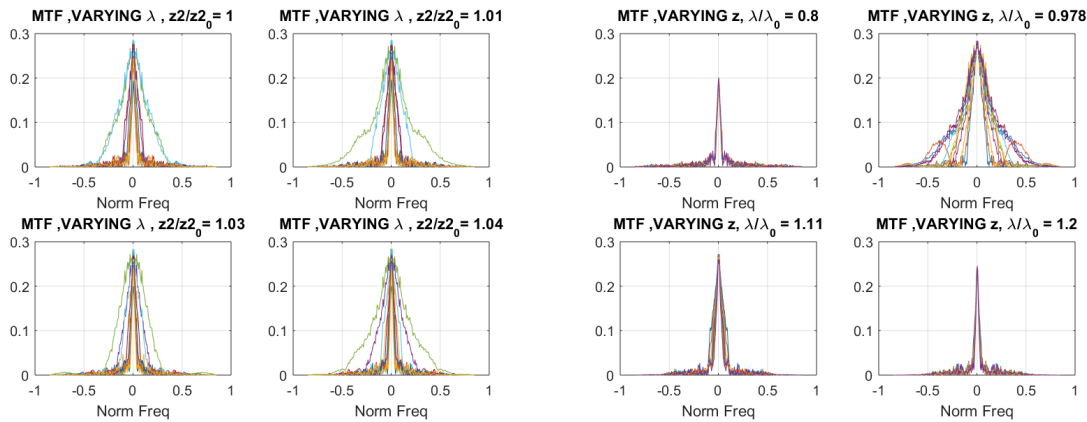

Fig. 12. The dependencies of MTF for the lensless system without the cubic phase term on variations in the defocus and the wavelength.

where $I_{o,\lambda}(u', v')$ is the object intensity radiation for the wavelength $\lambda$.

It is assumed in our modeling that

$$I_{o,\lambda}(u', v') = \gamma_\lambda I_o(u', v'), \quad (15)$$

i.e. there is an object pattern $I_o(u', v')$ independent on the wavelength $\lambda$ and the intensity of radiation is identical for all $\lambda$ within a scalar factor $\gamma_\lambda$.

The imaging problem is reformulated as a reconstruction of this wavelength independent object pattern $I_o$. Accordingly the observation model (14) is replaced by the weighted mean

$$I_s(u,v) = \frac{1}{\sum_\lambda \gamma_\lambda} \sum_\lambda \iint_{-\infty}^{\infty} \gamma_\lambda PSF_{\lambda,\lambda_0}(u-u', v-v') \ I_o(u',v')du'dv'. \quad (16)$$

In our experiments, the weights $\gamma_\lambda$ are selected following the spectral distribution of the black body radiation, which depends on the object temperature.

The summation (or integration) on $\lambda$ in (16) gives the total intensity of the radiation registered by the sensor.

Defining the averaged PSF as

$$\widehat{PSF}_{\lambda_0}(u,v) = \frac{1}{\sum_\lambda \gamma_\lambda} \sum_\lambda \gamma_\lambda PSF_{\lambda,\lambda_0}(u,v), \quad (17)$$

we arrive to the initial form of the input-output modeling (5), where $PSF$ is replaced by the averaged $\widehat{PSF}_{\lambda_0}(u,v)$ calculated for a given waveband and $I_{o,\lambda}$ is replaced by $I_o$ from (15).

In our modeling we assume that the distribution of $\gamma_\lambda$ is calculated accordingly to the Plank's law for the blackbody intensity radiation. The wavelength distribution of this thermal radiation is in the form

$$\tilde{\gamma}_\lambda = \frac{1}{\lambda^5} \exp(\frac{A}{\lambda T} - 1), \quad (18)$$
$$A = hc/k,$$

where $k = 1.38 \cdot 10^{-23}$ is the Boltzman constant; $c = 3*10^8$ is the light speed; $h = 6.6 \cdot 10^{-34}$ is the Plank constant.

All these constants are given in the corresponding units. Thus, the wavelength $\lambda$ in (18) is in $m$ and the absolute temperature $T$ is in Kelvin's degrees. In our simulation experiments $T = 300$ $K$. The $\gamma_\lambda$ in (15) is the normalized version of $\tilde{\gamma}_\lambda$: $\gamma_\lambda = \tilde{\gamma}_\lambda / \int_{\lambda_{low}}^{\lambda_{high}} \tilde{\gamma}_\lambda d\lambda$.

# References

[1] G. J. Swanson, "Binary optics technology: the theory and design of multi-level diffractive optical elements," , MIT, Lincoln Laboratory (1989).

[2] T. Grulois, G. Druart, N. Guérineau, A. Crastes, H. Sauer, and P. Chavel, "Extra-thin infrared camera for low-cost surveillance applications," Opt. Lett. 39, 3169-3172 (2014).

[3] E. R. Dowski and Cathey W. T. "Extended depth of field through wavefront coding," Appl. Opt., Vol. 34, No. 11, pp. 1859 – 1866 (1995).

[4] W. T. Cathey and E. R. Dowski, "New paradigm for imaging systems," Appl. Opt. 41, 6080-6092 (2002).

[5] G. Muyo, A. Singh, M. Andersson, D. Huckridge, A. Wood, and A. R. Harvey, "Infrared imaging with a wavefront-coded singlet lens," Opt. Express 17, 21118-21123 (2009).

[6] E. R. Dowski, "Wave front coded imaging systems," No.: US007554731B2, Jun. 30 (2009).

[7] S. Chen, Z. Fan, Z. Xu, B. Zuo, S. Wang, and H. Xiao, "Wavefront coding technique for controlling thermal defocus aberration in an infrared imaging system," Optics Letters, Vol. 36, Issue 16, pp. 3021-3023 (2011).

[8] M. Demenikov, E. Findlay and A. R. Harvey, "Miniaturization of zoom lenses with a single moving element," Optics Express, vol. 17, no. 8, pp. 6118-6127 (2009).

[9] M. Demenikov, "Development of compact optical zoom lenses with extended depth-of-field," Proc. SPIE 8488, Zoom Lenses IV, 84880C (2012).

[10] M. Demenikov, E. Findlay, A. R. Harvey, "Experimental demonstration of hybrid imaging for miniaturization of an optical zoom lens with a single moving element," Optics Letters 36(6), pp. 969-971 (2011).

[11] E. R. Dowski, I. Prischepa, "Wavefront coding zoom lens imaging system," No.: US 2003/0057353 A1 (2003).

[12] E. Findlay, A. Harvey, M. Demenikov,"Compact optical zoom," No.: US 008203627B2 (2012).

[13] S. Sherif, T. Cathey, and E. Dowski, "Phase plate to extend the depth of field of incoherent hybrid imaging systems," Appl. Opt. 43, 2709–2721 (2004).

[14] Q. Yang, L. Liu, and J. Sun, "Optimized phase pupil masks for extended depth of field," Opt. Commun. 272, 56–66 (2007).

[15] L. Ledesma-Carrillo, R. Guzmán-Cabrera, C. M. Gómez-Sarabia, M. Torres-Cisneros, and J. Ojeda-Castañeda, "Tunable field depth: hyperbolic optical masks," Applied Optics Vol. 56, Issue 1, pp. A104-A114 (2017)

[16] V. Nhu Le, Z. Fan, N. Pham Minh, and S. Chena, "Optimized square-root phase mask to generate defocus-invariant modulation transfer function in hybrid imaging systems," Optical Engineering 54(3), 035103 (March 2015).

[17] F. Diaz, F. Goudail, B. Loiseaux, and Jean-Pierre Huignard, "Comparison of depth-of-focus-enhancing pupil masks based on a signal-to-noise-ratio criterion after deconvolution," JOSA A Vol. 27, Issue 10, pp. 2123-2131 (2010).

[18] V. Le, Z. Fan , S. Chen, D. Duong Quoc, "Optimization of wavefront coding imaging system based on the phase transfer function," Optik 127, 1148–1152 (2016).

[19] A. Wood, N. Bustin and G. Muyo, "Computational imaging systems for the infrared: revolution or evolution", OECD Conference center, Paris, (2012).

[20] J. Ojeda-Castañeda and C. M. Gómez-Sarabia, "Tuning field depth at high resolution by pupil engineering," Advances in Optics and Photonics Vol. 7, Issue 4, pp. 814-880 (2015).

[21] T. Vettenburg, "Optimal Design of Hybrid Optical Digital Imaging Systems," Ph.D. Dissertation, Heriot-Watt Univ. (2010).

[22] S. V. King, A. Doblas, N Patwary, G. Saavedra, M. Matrinez-Corral, and C. Preza,"Spatial light modulator phase mask implementation of wavefront encoded 3D computational-optical microscopy," Vol. 54, No. 29, Applied Optics 8587- 8595 (2015).

[23] G. Carles, G. Muyo, S. Bosch and A.R. Harvey, "Use of a spatial light modulator as an adaptable phase mask for wavefront coding," Journal of Modern Optics," Vol. 57, No. 10, 893–900 (2010).

[24] V. A. Soifer (Editor), "Methods for Computer Design of Diffractive Optical Elements," Wiley Series in Lasers and Applications, 2002.

[25] E. Ben-Eliezer, N. Konforti, B. Milgrom and E. Marom, "An optimal binary amplitude-phase mask for hybrid imaging systems that exhibit high resolution and extended depth of field," Optics Express, Vol. 16, no. 25, 20540-20561 (2008).

[26] S. Elmalem and E. Maron, "Infrared imaging-passive thermal compensation via simple phase mask," Romanian Reports in Physics, Vol. 65, No. 3, P. 700–710, 2013.

[27] X. Wan, B. Shen, and R. Menon, "Diffractive lens design for optimized focusing," J. Opt. Soc. Am. A , Vol. 31, No. 12 (2014).

[28] R. N. Zahreddine and C. J. Cogswell, "Total variation regularized deconvolution for extended depth of field microscopy, Appl. Opt. 54(9) 2244-2254 (2015)

[29] R. C. Gonzalez and R. E. Woods, Digital Image Processing, 3rd ed. (Prentice–Hall, 2006).

[30] K. Dabov, A. Foi, V. Katkovnik, and K. Egiazarian, "Image restoration by sparse 3d transform-domain collaborative filtering," in Proc. SPIE Electron. Imag., San Jose, CA, Jan. 2008, vol. 6812, p. 681 207.

[31] K. Dabov, A. Foi, V. Katkovnik, and K. Egiazarian, "Image denoising by sparse 3D transform-domain collaborative filtering", *IEEE Trans. Image Process.*, vol. 16, no. 8, 2080-2095 (2007).

[32] A. Danielyan, V. Katkovnik, and K. Egiazarian, "BM3D frames and variational image deblurring, " *IEEE Trans. Image Process.*, vol. 21, no 4, 1715 – 1728 (2012).

[33] F. Heide, M. Steinberger, Y.-T. Tsai, M. Rouf, D. Pajak, D. Reddy, O. Gallo, J. Liu, W. Heidrich, K. Egiazarian, et al., "Flexisp: A flexible camera image processing framework," *ACM Transactions on Graphics (TOG)*, vol. 33, no. 6, p. 231, 2014.

[34] J.W. Goodman, Introduction to Fourier Optics, MaGraw-Hill, 3nd edition, 2005.

[35] G. Carles , A. Carnicer, S. Bosch,"Phase mask selection in wavefront coding systems: A design approach," Optics and Lasers in Engineering 48, 779–785 (29) (2010).